\documentclass[10pt,a4paper]{llncs}
\usepackage[utf8]{inputenc}
\usepackage[T1]{fontenc}
\usepackage{graphicx}
\usepackage{longtable}
\usepackage{amsmath}
\usepackage{xspace}

\newcommand{\E}{\ensuremath{\mathcal{E}}}
\newcommand{\etal}{\emph{et al.\xspace}}
\newenvironment{myproof}{\begin{proof}}{\hfill\qed\end{proof}}
\usepackage{clrscode}
\title{Evacuation of rectilinear polygons\thanks{This research was
funded by the German Ministry for Education and Research
(BMBF) under grant number 03NAPI4 ``ADVEST'', which fully funded Chris Gray.}}
\author{S\'andor Fekete \and Chris Gray
\and Alexander Kr\"oller
}
\institute{
Department of Computer Science, TU Braunschweig, Germany.
{\tt\(\{\)fekete,gray,kroeller\(\}\)@ibr.cs.tu-bs.de}}
\index{Fekete, S\'andor}
\index{Gray, Chris}
\index{Kr\"oller, Alexander}
\date{}

\begin{document}

\maketitle

\begin{abstract}
We investigate the problem of creating fast evacuation plans for
buildings that are modeled as grid polygons, possibly containing exponentially
many cells. We study this problem in
two contexts: the ``confluent'' context in which the routes to exits 
remain fixed over time,
and the ``non-confluent'' context in which routes may change. 
Confluent evacuation plans are simpler to carry out, as they allocate
contiguous regions to exits; non-confluent allocation
can possibly create faster evacuation plans.  We give results on the
hardness of creating the evacuation plans and strongly polynomial 
algorithms for finding
confluent evacuation plans when the building has two
exits.  We also give a pseudo-polynomial time algorithm for
non-confluent evacuation plans. Finally, we show that the 
worst-case bound between confluent and non-confluent plans is $2-O(\frac{1}{k})$.
\end{abstract}

\section{Introduction}
\label{sec-1}

  A proper evacuation plan is an important requirement for the health
  and safety of all people inside a building.  When we optimize
  evacuation plans, our goal is to allow people to exit the building
  as quickly as possible.  In the best case, each of a building's
  exits would serve an equal number of the building's inhabitants.
  However, there might be cases in which this can not happen.  For
  instance, when there is a bottleneck between two exits, it might
  make sense for most of the building's inhabitants to stay on the
  side of the bottleneck closer to where they begin, even if this
  means that one exit is used more than the other.

  In this research, we study the computation of evacuation plans for
  buildings that are modeled as grid polygons.  We make the assumption
  that every grid square is occupied by exactly one person and that at
  most one person can occupy a grid square at any given time.  The first
  assumption may seem a bit contrived, but in many cases it is impossible
  for a building designer to know exactly where people will be in the
  building in the moments before an evacuation, and this pessimistic
  view of the situation is the only sensible one to take.  Also
  remember that in some cases, such as in airplanes, the situation
  in which nearly every bit of floor space is occupied before an
  evacuation is more common than the alternative.

  Evacuation plans can be divided into two distinct types.  In the
  first, signs are posted that direct every person passing them to a
  specific exit.  In the second, every person is assigned to a
  distinct exit that does not necessarily depend on the exits to which
  his or her neighbors are assigned.  The first type of evacuation
  plan generates what is known as a \emph{confluent flow} and the second
  generates a \emph{non-confluent flow}.  More precise definitions of these
  terms will be given later.

  It is clear that in a non-confluent flow, people can be evacuated as
  quickly as in a confluent flow, and we show that in certain
  instances people can be evacuated significantly more quickly in a
  non-confluent flow than in a confluent flow.  However, a confluent
  flow is significantly easier to carry out than a non-confluent one,
  so we also give results related to it.  We first show that the
  problem of finding an optimal confluent flow belongs to the class of
  NP-complete problems if the polygon has ``holes''---that is, if it
  represents a building with completely enclosed rooms or other
  spaces.  We then give an algorithm with a running time linear in the description
  complexity of the region (which can be exponentially smaller than
  the number of cells) that computes an evacuation plan for buildings
  without holes that have two exits; a generalization to a constant number
  of exits is more complicated, but seems plausible.  Finally, we show
  that the worst-case ratio between the evacuation times for confluent
  flows and non-confluent flows for $k$ exits is $2 - O(\frac{1}{k})$.

\subsection{Preliminaries}
\label{sec-1.1}

   We are given a rectilinear polygon \(P\) on a grid.  There exists,
   on the boundary of \(P\), a number of special grid squares known as
   \emph{exits}.  We call the set of exits \(\E = \{e_1, \ldots, e_k\}\).
   We assume that every grid square in \(P\) contains a person.  A
   person can move vertically or horizontally into an empty grid
   square or an exit.  The goal is to get each person to an exit as
   quickly as possible.  When an exit borders more than one grid
   square, we specify the squares from which people can move into the exit.

   The area of \(P\) is denoted by \(A\) and the number of vertices of
   \(P\) is denoted by \(n\).  Note that \(A\) can be exponential in
   \(n\).
   The set of people that leave \(P\) through the exit \(e\) is called
   the \emph{\(e\)-exit class}.  We also write \emph{exit class} to refer to the
   set of people who leave through an unspecified exit.
   The grid squares that are adjacent to the boundary of \(P\) are
   known as \emph{boundary squares}.

   There are two versions of the problem that we consider.  We call
   these \emph{confluent flows} and \emph{non-confluent flows}.  In the first,
   we add the restriction that every grid square has a unique
   successor.
   Thus, for every grid square \(s\), people passing through \(s\)
   leave \(s\) in only one direction.
   This restriction implies that evacuation plans are determined by space only.
   It does not exist for
   non-confluent flows.  It can be argued that informing
   people which exit to use is easier in the case of confluent flows
   since a sign can be placed in every grid square, informing the
   people who pass through it which exit to use.  However, we show
   that non-confluent flows can lead to significantly faster
   evacuations.

   The major difficulty in the problem comes from bottlenecks.  We
   define a \emph{\(k\)-bottleneck} to be a rectangular subpolygon \(B\) of
   \(P\) such that two parallel boundary edges of \(B\) are the same as two
   edges of \(P\), where the distance between the common edges is
   \(k\).  

\subsection{Related Work}
\label{sec-1.3}

   The problem when restricted to confluent flows is similar in many
   ways to the unweighted Bounded Connected Partition (or 1-BCP)
   problem~\cite{salgado_approximation_2004}.  In this problem,
   one is given an unweighted, undirected graph \(G = (V, E)\) and
   \(k\) distinguished vertices.  The goal is to find \(k\) connected
   subsets of \(V\), where each subset contains exactly one of the
   distinguished vertices and where each has the same cardinality.  If
   we define the dual of the grid contained in our polygon to be the
   graph formed by connecting adjacent faces and let the distinguished
   vertices be the exits, then 1-BCP is clearly the same as evacuation
   restricted to confluent flows with \(k\) exits.

   It was independently shown by
   Lovász~\cite{lovsz_homology_1977} and
   Györi~\cite{gyri_division_1978} that a solution to the 1-BCP
   problem can be found for every graph that is \(k\)-connected.
   However, their proofs are not algorithmic.
   Thus, there has been some work on finding partitions
   of size \(k\) for low values of \(k\).
For example, Suzuki \etal
give an algorithm for 2-partitioning 2-connected
graphs~\cite{suzuki_linear_1990}. 
One algorithm
   claiming a general solution for \(k\)-partitioning
   \(k\)-connected graphs~\cite{ma_$ok^2n^2$_1994} 
   is incorrect.

   Unfortunately, when \(P\) contains 1-bottlenecks, the graph
   obtained by finding the dual of the grid inside \(P\) is
   1-connected.  Therefore, the results of Lovás and Györi do not
   apply.  Also, since the dual of the grid contained in
   our polygon can have size exponential in the complexity of the
   polygon, we would probably need to merge nodes and then assign
   weights to the nodes of the newly-constructed graph.  The addition
   of weights, however, makes the BCP problem NP-complete, even in the
   restricted case in which the graph is a
   grid~\cite{becker_max-min_1998}.

   Another connection is to the problem of partitioning polygons into
   subpolygons that all have equal area.  The confluent version of our
   problem, indeed, can be seen as the ``discrete'' version of that
   problem.  The continuous version has been studied.  One interesting
   result from this study is that finding such a decomposition while
   minimizing the lengths of the segments that do the partitioning is
   NP-hard even when the polygons are
   orthogonal~\cite{bast_area_2000}.  However, polynomial
   algorithms exist for the continuous case when that restriction is
   removed~\cite{lumelsky_polygon_1998}. 

   Baumann and Skutella~\cite{bs-eafms-09} consider evacuation problems modeled
   as earliest-arrival flows with multiple sources. 
   They achieved a
   strongly-polynomial-time algorithm by showing that the function
   representing the number of people evacuated by a given time is
   submodular.  Such a function can be optimized using the parametric
   search technique of Megiddo.  Their approach is different from ours in that
   they are given an explicit representation of the flow network as input.
   We are not given this, and computing the flow network that is implicit
   in our input can take exponential time.  Also, their algorithm takes
   polynomial time in the sum of the input and output sizes.  However, the
   complexity of the output can be exponential in the input size.

   Another, perhaps more surprising, related problem is machine
   scheduling.  
Viewed simply, confluent flows correspond to
   non-preemptive scheduling problems, while non-confluent
   flows correspond to preemptive scheduling problems. 
The
   NP-hardness result for optimal confluent flows is inspired by the hardness of scheduling jobs on
   non-preemptive machines~\cite{garey_computers_1979}, while
the worst-case ratio between confluent and non-confluent flows
   is inspired by the list-scheduling approximation ratio~\cite{graham_boundsmultiprocessing_1969}.
   
\section{Confluent Flows}
\label{sec-2}

  \label{sec:confluent}

  As mentioned in the introduction, 
  in a confluent flow, every grid square has the property that all
  people that pass through it use the same exit.

  In this section, we present our results related to confluent flows.
  First, we show the NP-completeness of the problem of finding an
  optimal evacuation plan with confluent flows in a polygon with
  holes.  This holds even for polygons with two exits.  We then give
  a linear-time algorithm for polygons with two exits.
  
\subsection{Hardness}
\label{sec-2.1}

   \label{sec:hardness}

   \paragraph{Weak NP-hardness with two exits.}
   We first show that the evacuation problem with confluent flows is
   NP-hard if we allow \(P\) to have holes.  We reduce from the
   problem \textsc{Partition}, which is well-known to be
   NP-complete~\cite{garey_computers_1979}.  In this problem, we
   are given a set \(S = \{c_1, c_2, \ldots, c_m\}\) of integers and
   we are asked to determine whether we can find \(S_1, S_2 \subseteq S\)
   such that \(\sum_{c_h \in S_1} c_h =
   \sum_{c_i \in S_2} c_i\).

   We note that if we scale all of the numbers in a \textsc{Partition}
   instance by an integer \(\ell\), the answer remains the same---that
   is, a partition can be found in the new set if and only if a
   partition could be found in the old set---but the difference
   between the size of non-optimal sets is at least \(\ell\).  This is
   because \[\sum_{c_h \in S_1} \ell c_h - \sum_{c_i \in S_2} \ell c_i =
   \ell\left(\sum_{c_h \in S_1} c_h - \sum_{c_i \in S_2} c_i\right) \ge
   \ell\] if the sums are not equal.

\begin{figure}[tb]
\centerline{\includegraphics[width=20em]{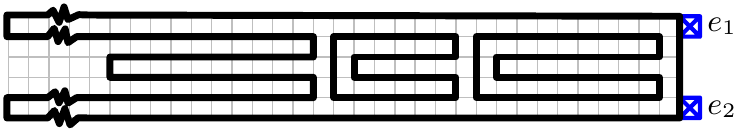}}
\caption{\label{fig:np-hardness}The polygon \(P\) given a \textsc{Partition} instance of \(\{11, 6, 9\}\).  To keep the picture a manageable size, the elements have not been scaled and the left ends of the first and fifth rows are truncated.}
\end{figure}

   To transform the \textsc{Partition} problem into our problem, we
   first scale the input by a factor of \(2m + 1\).  We then do the
   following to make the polygon \(P\):

\begin{itemize}
\item We make a rectangle whose width is \(m + 1 + \sum_{c_i \in S} c_i\) and whose height is 5.
\item We remove all the grid squares of \(P\) on the second and fourth
     rows except those that are at position \(\sum_{i = 1}^j c_i +
     j\) for all \(0 < j \le m\).
\item We remove all grid squares from the third row that are at
     position \(\sum_{i = 1}^j c_i + j\) for all \(0 < j \le m\).
\item We add a large number of squares (at least equal to the current area of \(P\)) to
     the left end of the first and fifth rows.
\item We add an exit \(e_1\) to the right end of the first row and an exit
     \(e_2\) to the right end of the fifth row.
\end{itemize}
   See Figure \ref{fig:np-hardness} for a small example.  We say that
   the connected sets of grid squares in the third row each correspond
   to one of the elements of the given \textsc{Partition} instance.
   This leads to the following lemma.

\begin{lemma}
   \label{th:confluent-with-holes-equiv-to-partition}
   The polygon \(P\) with holes can be divided into two confluent exit
   classes of equal size if and only if the given instance of
   \textsc{Partition} has a ``yes'' answer.
\end{lemma}

\begin{myproof}
   First, assume that the \textsc{Partition} instance has a ``yes''
   answer and that we are given \(S_1\) and \(S_2\), the sum of whose
   elements are equal.  We then take the grid squares from the third
   row that correspond to elements in the set \(S_1\) and send them to
   \(e_1\).  We take the remaining grid squares from the third row and
   send them to \(e_2\).  We also send all the grid squares in the
   first and second rows to \(e_1\) and all the grid squares in the
   fourth and fifth rows to \(e_2\).

   It is obvious that the number of grid squares in the rows other
   than the third row are equal.  Since the sum of the elements in
   \(S_1\) is equal to the sum of the elements in \(S_2\), the number
   of grid squares sent to \(e_1\) and \(e_2\) is equal.

   Now assume that the \textsc{Partition} instance does not have a
   ``yes'' answer.  This implies that \(|\sum_{c_i \in S_1} c_i -
   \sum_{c_j \in S_2} c_j| \ge 2m + 1\) for any \(S_1\) and \(S_2\)
   that are subsets of \(S\).  Since we are dealing with a confluent
   flow, if any grid square from the first row exits through \(e_2\),
   then all grid squares to the left of it must exit through \(e_2\).  This would
   mean that the number of grid squares exiting through \(e_2\) would be
   more than \(2/3\) the number of grid squares in \(P\) (because of the
   ``large number'' of grid squares added to the left end of the first and
   fifth rows).  Therefore, we can assume that no grid square from the
   first row exits through \(e_2\) and similarly we can assume that
   no grid square from the fifth row exits through \(e_1\).  This implies
   that the only grid squares that can ``choose'' which exit to go through
   are those in the second, third, and fourth rows.

   There are exactly \(2m\) grid squares in the second and fourth rows.
   Because the \textsc{Partition} instance does not have a ``yes''
   answer, the number of grid squares from the third row going through
   \(e_1\) has a difference of at least \(2m + 1\) with the number of
   grid squares going through \(e_2\).  Therefore, there is no way that the
   number of grid squares going through \(e_1\) is equal to the number of
   grid squares going through \(e_2\).
\end{myproof}

   We define the decision version of the problem of evacuation with
   confluent flows to be: ``Given a grid polygon \(P\) with \(k\) exits
   and a natural number \(\ell\), can a confluent flow be found in which  
   the largest exit class has size at most \(\ell\)?''

   Since areas of polygons can be computed in time proportional to
   their number of vertices, we can verify if a solution is correct in
   \(O(kn)\) time (where \(n\) is the number of vertices of the
   polygon and \(k\) is the number of exits).  This, along with Lemma
   \ref{th:confluent-with-holes-equiv-to-partition}, implies that the
   decision problem is NP-complete in polygons with holes.  We
   summarize this result in the following theorem.

\begin{theorem}
   \label{th:weak-np-complete}
   The problem of finding an optimal confluent flow in a polygon with
   holes is NP-complete.
\end{theorem}

   \paragraph{Strong NP-hardness.}
   Theorem~\ref{th:weak-np-complete} shows that the problem is
   \emph{weakly NP-complete}.  This means that the hardness of the problem
   depends on the areas of subpolygons being exponential in the
   complexity of the input.  This implies that a pseudo-polynomial
   algorithm might exist.

   However, we will now show that if we allow \(O(n)\) exits, the
   problem is \emph{strongly NP-complete}.  This means that the problem is
   still NP-complete when all of its numerical parameters are
   polynomially bounded in the size of the input.

   Our reduction is from \textsc{Cubic Planar Monotone 1-in-3
   Satisfiability} (or \textsc{CPM 1-in-3 SAT} for short).  This is a
   variant of the \textsc{Satisfiability} problem in which
\begin{itemize}
\item every clause contains exactly 3 literals,
\item every variable is in exactly 3 clauses,
\item the graph generated by the connections of variables to clauses is
     planar,
\item every literal in every clause is non-negated, and
\item a clause is satisfied if exactly one of its variables is true.
\end{itemize}
   We use a reduction that is almost equivalent to the one showing
   that tiling a finite subset of the plane with right trominoes is
   NP-complete~\cite{moore_hard_2001}, so we will summarize that
   reduction and describe our modifications to it.

   A \emph{right tromino} is a type of tile that consists of only three grid
   squares arranged in an ``L'' shape.  Moore and Robson
   showed~\cite{moore_hard_2001} that tiling a finite subset of
   the plane (that is, a polygon with holes) with right trominoes
   is NP-complete.  Since their reduction contains no numerical
   parameters and all the coordinates used are polynomially-sized
   integers, it is clear that the problem is strongly NP-complete.

   The reduction uses gadgets---small sections of polygons that can be
   put together to form a larger polygon---to create a polygon that
   can be tiled by right trominoes if and only if a given instance of
   \textsc{CPM 1-in-3 SAT} is satisfiable.  The reduction is as
   follows.  First, the graph \(G = (V, E)\) generated by connecting
   variables to clauses is
   embedded on a grid.  The embedding must have the property that
   every vertex is on a grid point and the set of edges is a set of
   disjoint paths along grid lines.  Then every grid point that is
   occupied is replaced by an appropriate gadget (rotated
   appropriately).  The vertices that correspond to variables are
   replaced by variable gadget; those that correspond to clauses are
   replaced by a clause gadget.  A grid point that is occupied
   by an edge is replaced by a wire gadget when the edge continues in
   the same direction through the grid point and by a turn gadget when
   the edge turns at the grid point.  Since \(G\) is a planar
   graph, it should be clear that the constructed polygon has holes.

\begin{figure}[tb]
\centerline{\includegraphics[width=.6\columnwidth]{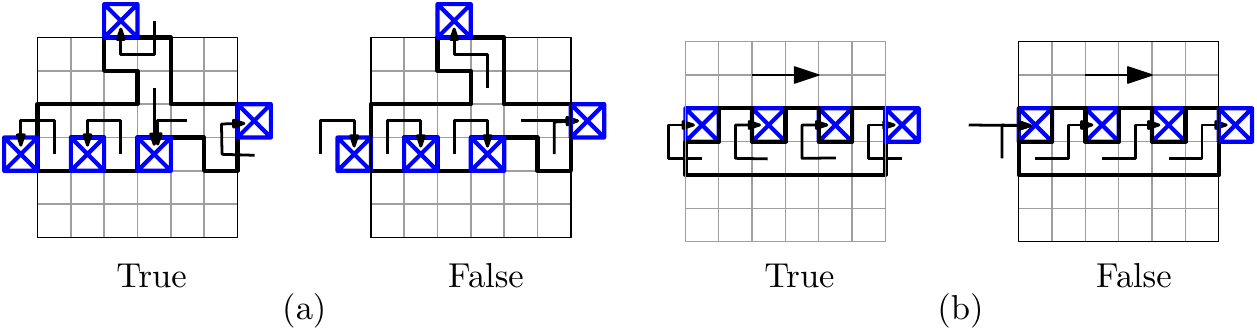}}
\caption{\label{fig:variable-gadget}\label{fig:wire-gadget}(a) The variable gadget. (b) The
wire gadget.}
\end{figure}

\begin{figure}[tb]
\centerline{\includegraphics[width=.6\columnwidth]{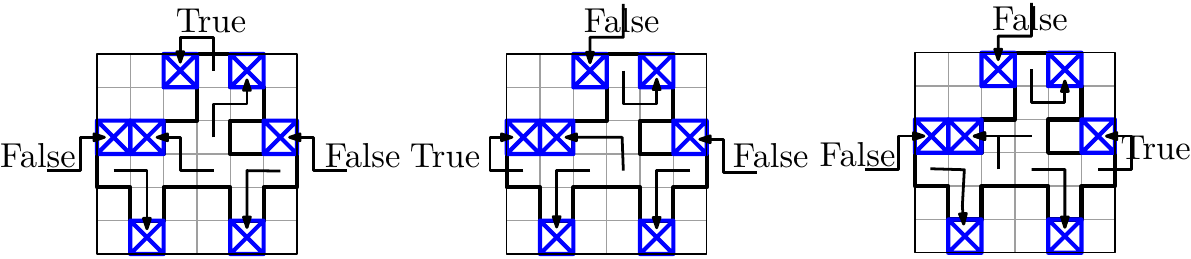}}
\caption{\label{fig:clause-gadget}The clause gadget, showing the three
ways that people can evacuate such that each exit handles exactly
three people.}
\end{figure}

\begin{figure}[tb]
\centerline{\includegraphics[width=.3\columnwidth]{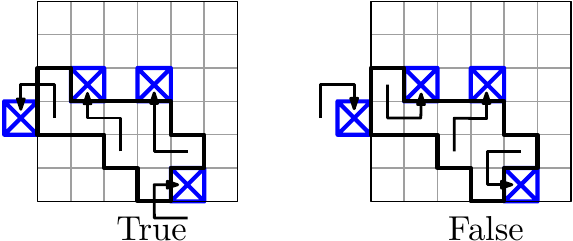}}
\caption{\label{fig:turn-gadget}The turn gadget.}
\end{figure}

   The gadgets shown in
   Figures~\ref{fig:variable-gadget}--\ref{fig:turn-gadget} are the
   same as those in the proof by Moore and Robson, we have only added
   exits to them so that the entire polygon can be evacuated with a
   maximum of three people going through each exit if and only if the
   polygon can be tiled by right trominoes.   

   The paths indicated in
   Figures~\ref{fig:variable-gadget}--\ref{fig:turn-gadget}
   indicate the possible evacuation routes that satisfy the
   requirement that every exit handles at most three people.  They
   correspond to the possible tiling of each gadget by right trominoes
   in the proof by Moore and Robson.  
We summarize.
   
\begin{theorem}
   The problem of finding an optimal confluent flow
   in a polygon with holes and \(O(n)\) exits is strongly NP-complete.
\end{theorem}
   
   Since this implies that finding an optimal confluent flow in a polygon with
   holes in polynomial time is unlikely, we assume in the rest of the
   paper that \(P\) is a simple grid polygon without holes.

\subsection{Two Exits}
\label{sec-2.3}

   When the polygon \(P\) has only two exits, \(e_1\) and \(e_2\), we
   can find an optimal confluent flow in \(O(n)\) time.  We first
   present an algorithm that takes cubic time that can be
   modified fairly easily into an algorithm that takes linear time. 
   
\subsubsection{Na\"ive algorithm}
\label{sec-2.3.1}

    \label{sec:naive}

    Notice that the case in which \(P\) has two exits is simpler than the
    case in which \(P\) has more exits because of the fact that both exit
    classes must each have one contiguous connection to the boundary of
    \(P\).  

    We begin with a decomposition of \(P\) into rectangles.  This
    decomposition is the overlay of two simpler decompositions: the
    vertical and horizontal decompositions.  The vertical
    decomposition of a rectilinear polygon \(P\) is the partition of
    \(P\) into rectangles by the addition of only vertical line
    segments.  Similarly, the horizontal decomposition of \(P\) is the
    partition of \(P\) into rectangles by the addition of only
    horizontal line segments.  We call the overlay \(\omega\) and its
    dual graph \(\omega^*\).  We add the vertical and horizontal line
    segments from the grid points on opposite sides of \(e_1\) and
    \(e_2\) to \(\omega\) as well.  We then do the following for every
    pair of rectangles \(r_1\) and \(r_2\) in \(\omega\) that have at
    least one edge of the boundary of \(P\).  We first ensure that
    \(e_1\) is between \(r_1\) and \(r_2\) and that \(e_2\) is between
    \(r_2\) and \(r_1\).  We also ensure that there are no bottlenecks
    of size 1 between \(r_1\) and \(r_2\).  If either of these
    conditions are not met, we proceed to the next pair of rectangles.
    We then set the \(e_1\)-exit class to be all the grid squares
    along the boundary of \(P\) between \(r_1\) and \(r_2\).  We then
    add all the grid squares surrounded by the \(e_1\)-exit class to
    the \(e_1\)-exit class.  We call the area of the \(e_1\)-exit
    class \(A_1\).  We define the \(e_2\)-exit class similarly and
    call its area \(A_2\).  We call the larger of the two areas
    \(A_\ell\) and its corresponding exit \(e_\ell\). If \(A_\ell\) is
    less than \(A / 2\), we can divide the rest of the grid squares
    evenly among the exit classes---see Lemma~\ref{lem:connected} for details---and return the solution.  Otherwise,
    we attempt to make the \(e_\ell\)-exit class as small as possible
    (while staying above \(A / 2\)) inside \(r_1\) and \(r_2\) and
    assign the rest of the grid squares to the other exit class.  We
    maintain a variable that tracks the area of the
    smallest such exit class \(e_{min}\).  If we get through all the
    possible pairs \(r_1\) and \(r_2\) without finding a pair that
    we can return, we return~\(e_{min}\). 

\begin{lemma}
\label{lem:naive-correct}
  The algorithm presented above is correct and takes \(O(n^3)\) time.
\end{lemma}

\begin{myproof}
  We assume that the algorithm presented in Section~\ref{sec:connected}
  is correct and runs in \(O(n)\) time.

  The time complexity of the algorithm comes from the fact that there
  are \(O(n)\) rectangles of \(\omega\) with at least one edge on the
  boundary of the polygon.  We loop through each pair of these
  rectangles, and in each loop there is the possibility that we must
  find the area of an exit class.

  To show that the algorithm is correct, we must argue that it finds
  the \(e_1\)- and \(e_2\)-exit classes and that
\begin{itemize}
\item both classes are connected and
\item the size of the larger class is as small as possible.
\end{itemize}
Clearly, if we find connected exit classes that both have size \(A /
2\) (which we do if we return early), we have satisfied both
requirements.  We find such a solution if it exists because we try all
combinatorially unique starting and ending points for the connection
of the \(e_1\)-exit class to the boundary of \(P\).

  Therefore, it remains to argue that we find the optimum solution for
  the case in which one exit class is larger than \(A / 2\).  In this
  case, we also try all combinatorially unique starting and ending
  points that minimize the size of both the \(e_1\) and \(e_2\) exit
  classes.  The exit classes that we return consist of the smallest
  exit class that is larger than \(A / 2\) and its complement (which
  must be smaller).  This proves the lemma.
\end{myproof}

\subsubsection{Linear algorithm}
\label{sec-2.3.2}

The algorithm above has two steps that lead to it taking cubic time:
the loop over all pairs of rectangles on the boundary of \(P\) and the
computation of the minimum area for an exit class that has a
connection to the boundary that begins in one of the rectangles and
ends in the other.  In this section, we give a more clever solution
that avoids these problems.

    We begin by observing that if we update the area, each time we
    change the starting and ending points of the connection of the
    \(e_1\)-exit class to the boundary of the polygon rather than
    computing it anew, the total time spent computing the area depends
    on the sum of the complexities of the updated areas.

    We also observe that we loop over the rectangles of \(\omega\)
    with at least one edge attached to the boundary of \(P\).  This
    means that we do not really need to compute the entire overlay
    \(\omega\)---only the intersections of \(\omega\) with the
    boundary of \(P\).  These intersections can be computed in
    \(O(n)\) time by computing the vertical and horizontal
    decompositions of \(P\) separately.  We call the set of intervals 
    thus computed \(\omega'\).  Once again, we are given the exits
    \(e_1\) and \(e_2\).  

    We create two pointers \(i_1\) and \(i_2\) with which we walk
    through the intervals in \(\omega'\).  For each pair of intervals
    pointed to by \(i_1\) and \(i_2\) that we visit, we measure the
    number of squares that must be in the \(e_1\)- and \(e_2\)-exit
    classes if we assume that the endpoints of their connections to
    the boundary of \(P\) begin and end in \(i_1\) and \(i_2\).  We
    call these areas \(A_1\) and \(A_2\) respectively.  If we ever
    visit a pair of intervals for which \(A_1\) and \(A_2\) are both less
    than \(A / 2\), then we divide the remaining squares so that both
    \(A_1\) and \(A_2\) are \(A / 2\)---see Lemma~\ref{lem:connected}
    for details---and return
    the results.  Otherwise, we return the exit class that has size
    greater than \(A / 2\), but whose size is minimal.


    \begin{figure}[htbp]
    \centering
    \includegraphics[width=0.75\columnwidth]{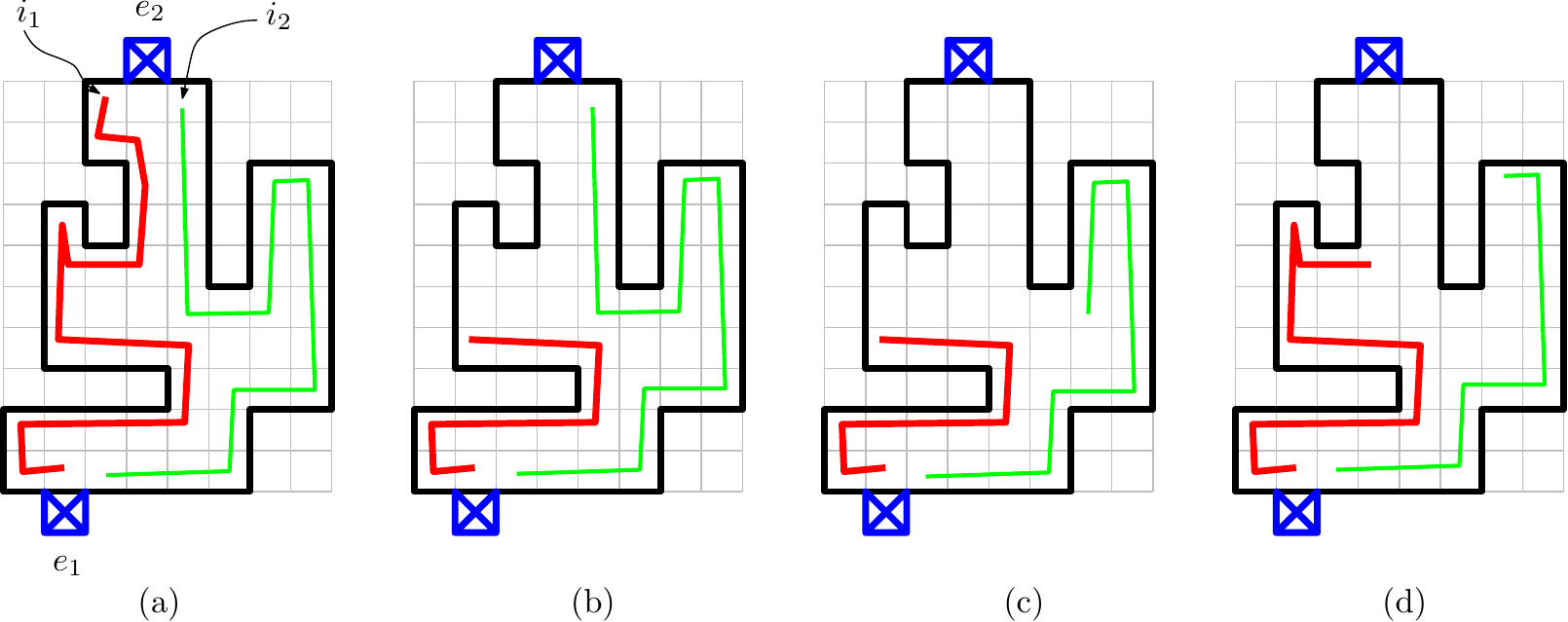}
    \caption{An example of the linear-time algorithm.  The pointers
    \(i_1\) and \(i_2\) denote the endpoints of the connection of the
    \(e_1\)-exit class to the boundary of the polygon.}
    \label{fig:linear-alg}
    \end{figure}

    To begin with, we set \(i_1\) and \(i_2\) to be the interval
    containing \(e_2\), so that the endpoints of the connection of the
    \(e_1\)-exit class are on either side of \(e_2\).  See Figure
    \ref{fig:linear-alg} (a).  
We call the area that the \(e_1\)-exit
    class must have \(A_1\) and the area that the \(e_2\)-exit class
    must have \(A_2\).  We then move \(i_1\) closer to \(e_1\) until
    it either reaches \(e_1\) or would cause \(A_2\) to be greater
    than \(A / 2\).  As we progress, we simply update \(A_1\) and
    \(A_2\) and keep track of the smallest value for \(A_1\).  See
    Figure \ref{fig:linear-alg} (b).

    Once we have done this for \(i_1\), we do the same for \(i_2\).
    See Figure \ref{fig:linear-alg} (c).  
Finally, we move \(i_1\)
    back towards \(e_2\).  For each interval that we move \(i_1\)
    towards \(e_2\), we move \(i_2\) as much as possible towards
    \(e_1\) so that \(A_1\) is as small as possible without causing
    \(A_2\) to be larger than \(A / 2\).  See Figure
    \ref{fig:linear-alg} (d).  
As before, we keep track of the
    smallest value for \(A_1\).  When \(i_1\) reaches \(e_2\) or
    \(i_2\) reaches \(e_1\), we stop.

    When we have completed the algorithm for \(e_1\), we repeat the process,
    switching \(e_1\) and \(e_2\).

\begin{theorem}
\label{th:confluent-linear}
    In the confluent setting, the above algorithm finds the optimal evacuation plan for a
    polygon \(P\) 
    with two exits in \(O(n)\) time, where \(n\) is the number of
    vertices in \(P\).
\end{theorem}

\begin{myproof}
    It is easy to see that this algorithm runs in linear time since
    each time we move a pointer, we only need to update the area
    currently assigned to the \(e_1\)-class.  This takes time
    proportional to the complexity of the area that we are adding or
    removing, and the sum of the complexities of the updated areas is
    at most \(2n\).

    We must argue, however, that the algorithm finds the minimum value
    for \(A_1\) if \(A_1\) must be greater than \(A / 2\) in the
    optimal layout (a symmetric argument holds for \(A_2\)).  The
    first two phases of the algorithm simply reduce \(A_1\) as much as
    possible while keeping \(i_2\) as close to \(e_2\) as possible.
    The third phase is similar to the \emph{rotating calipers}
    algorithm~\cite{toussaint_solving_1983}.  An invariant is
    established---in this case, \(A_1\) is as small as possible for
    the given interval \(i_2\) without making \(A_2\) be greater than
    \(A / 2\)---and then two pointers are moved in the same direction
    around \(P\) while maintaining the invariant.  In our algorithm
    \(i_2\) visits every reasonable interval between \(e_1\) and
    \(e_2\), meaning that we eventually find the optimum interval.
\end{myproof}

\subsection{Dividing the interior}
\label{sec:connected}

As noted in the algorithms in Section~\ref{sec:naive}, we need a subroutine that assigns
grid squares to exits once the grid squares along the boundary of
\(P\) have been assigned.  Given a connected subpolygon \(P' \subseteq
P\), a natural number \(x\), and an initial connected set \(E_i
\subseteq P'\) of squares that belong to the exit class \(e\) 
we wish to find a set \(S\)
of squares, where \(|S| = x\) (the \emph{size}
condition) and where both \(S \cup E_i\) and \(P - (S \cup E_i)\) are
connected (the \emph{connectivity} condition).  More precisely, since
\(x\) could be exponential in the complexity of the input, we would
like to find a polygon that has complexity linear in the input that
represents the boundaries of \(S\).

   \begin{figure}[htb]
\centerline{\includegraphics[width=20em]{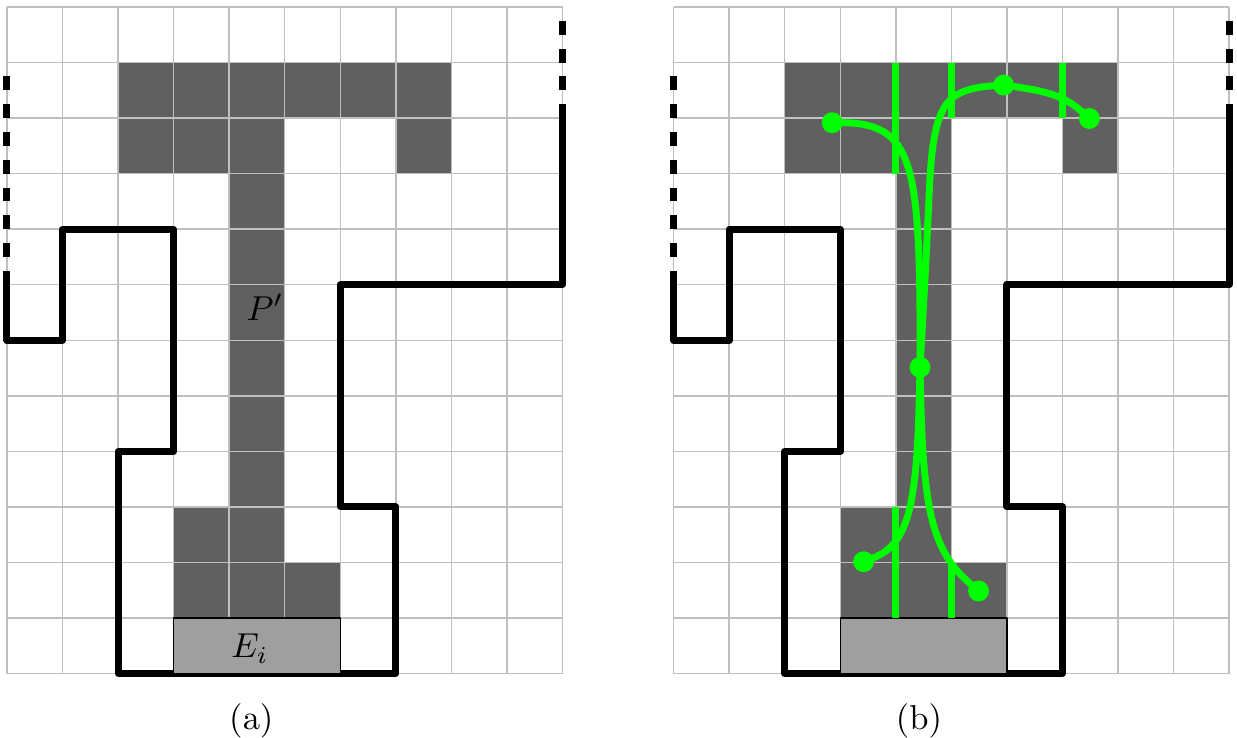}}
\caption{\label{fig:connected-traversal}(a) \(P'\) and \(E_i\).  (b) The vertical decomposition and its dual.}
\end{figure}

   We solve this problem for the useful case in which \(P'\) is
   assumed to have no boundary squares except those that are in
   \(E_i\).  In this case, we do not need to worry about the
   connectivity condition as long as \(S \cup E_i\) is a connected
   polygon without holes.  This is because any square in \(P - (S \cup
   E_i)\) has a path to the boundary.  Since \(E_i\) is a connected
   set, the boundary squares of \(P - (S \cup E_i)\) are a connected
   subset of \(P\).  

   It is thus relatively simple to find \(S\) given the assumption
   that \(P'\) may not contain any boundary squares of \(P\).  First,
   we check whether \(|P'| > x\).  If it is not, we simply return the
   whole of \(P' \cup E_i\).  Otherwise, we find the vertical
   decomposition of \(P' \cup E_i\).  We perform a
   breadth-first-traversal of the dual of this vertical
   decomposition 
starting at a
   rectangle that contains a square of \(E_i\), and adding the squares
   from each visited rectangle to \(S\) as we go.  We stop the
   traversal when \(|S| + |E_i| \ge x\).  We subtract the appropriate
   amount of squares from \(S\) and return it.

\begin{lemma}
\label{lem:connected}
   The above algorithm runs in \(O(n)\) time and has output size that
   is linear in the input size.
\end{lemma}

\begin{myproof}
   Finding a vertical decomposition of a polygon can be completed in
   linear time~\cite{chazelle_triangulatingsimple_1991}.  Since
   the area of each rectangle of the decomposition can be computed in
   constant time, the time required for the breadth-first-traversal of
   the dual of the decomposition is also \(O(n)\).  Finally, we note
   that each rectangle that we add during the traversal can be charged
   to one of the vertices of the input, proving the second part of the
   lemma.
\end{myproof}

We conjecture that one can use algorithms similar to the na\"ive
algorithm given above to compute the evacuation of any polygon with a
constant number of exits, but the details become much more involved.
We therefore leave this question to future work.
   
\section{Non-Confluent Flows}
\label{sec-3}

  \label{sec:non-confluent}

  Compared with confluent flows, non-confluent flows are clearly a
  stronger model.  We note that any confluent flow is a non-confluent
  flow, but not \emph{vice versa}.  We show that non-confluent flows can be
  as much as \(2 - 2 / (k + 1)\) times as fast as confluent flows by
  giving an example in which this is the case.  We then argue that the
  ratio our example achieves is optimal.

\subsection{Pseudo-Polynomial Algorithm}

In contrast to the case with confluent flows, for which we showed that
finding an assignment of people to exits is strongly NP-complete when
we are dealing with polygons with holes and \(O(n)\) exits, we can
show that, for non-confluent flows, a pseudo-polynomial algorithm
exists.

The algorithm is based on the technique of using time-expanded
networks to compute flows over time~\cite{skutella_introduction_2009}.
Therefore, we compute a flow network from the input polygon as
follows.  We create a source vertex \(s\) and a sink vertex \(t\).
For each grid square in \(P\), we create two vertices---an \emph{in}
vertex and an \emph{out} vertex.  We connect the in vertex to the out
vertex with an edge that has capacity 1 for every grid square.  We
then make, for some integer \(T \ge 1\), \(T\) copies of the polygon
\(P_1, \ldots, P_T\), where each copy has these vertices and edges
added.  For every grid square of \(P_1\), we connect \(s\) to the in
vertex of the grid square with an edge that has capacity 1.  We then
connect the out vertex of every grid square in \(P_i\) to the in
vertex of all its neighbors in \(P_{i + 1}\) for all \(1 \le i \le T -
1\).  Again, the edges we use all have capacity 1.  Finally, we
connect the out vertex of every exit to \(t\) with an edge that has
capacity 1.  We call this flow network \(G\).

It is fairly easy to see that if we are able to find a maximum flow of
value \(A\) through \(G\), then we are able to evacuate \(P\) in \(T\)
time steps.  However, we note that both \(T\) and \(|G|\) can be
exponential in the complexity of \(P\), making this a
pseudo-polynomial algorithm.

\begin{theorem}
There exists a pseudo-polynomial algorithm to find an evacuation of a
polygon with a non-confluent flow.
\end{theorem}

\subsection{Differences to Confluent Flows}

\begin{figure}[tb]
\begin{tabular}{c@{\hspace{5em}}c}
\includegraphics[]{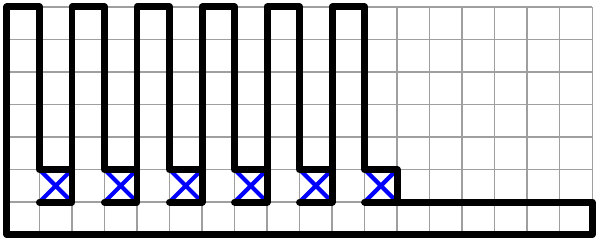} &
\includegraphics{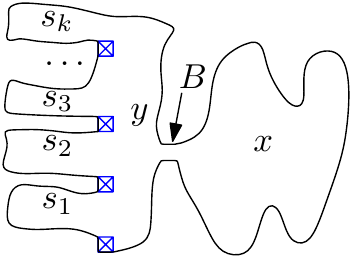} \\
(a) & (b) \\
\end{tabular}
\caption{\label{fig:non-confluent-example} (a) A polygon whose optimal
non-confluent flow is nearly twice as fast as its optimal confluent
flow. (b) \label{fig:max-ratio-shape} The general shape of any polygon
that realizes the maximum ratio between the confluent and
non-confluent flows.}
\end{figure}

  The example that shows a large gap between confluent and
  non-confluent flows is a horizontal rectangle of
  width 1 with length \(2k + mk\), for some integer \(m \ge 1\).  Attached to this rectangle are \(k\)
  vertical rectangles of width 1 and length \(mk\)---one at every other
  square for the first \(2k\) squares.  Between each vertical
  rectangle is an exit.  
  Each exit can only be entered from the square to the left.  See Figure \ref{fig:non-confluent-example}(a).

  We can see that the example has an optimal
  confluent flow that requires \(2mk + 3\) time steps: three to remove
  the people directly to the left of each exit and all the people
  below the exits, \(mk\) to remove the people in the vertical
  rectangles, and another \(mk\) for the people in the horizontal
  ``tail'' to go through the rightmost exit.  On the other hand, in the
  optimal non-confluent flow, all exits can remain continuously busy.
  One way that this can happen is for \(m\) people from the horizontal
  rectangle to
  leave through successive exits, while people from the vertical
  rectangles are leaving through the other exits.  Since the exits are
  continuously busy, the amount of time for all people to leave is
  \((k^2m + (2 + m)k) / k = mk + 2 + m\).  
  The ratio between the confluent and non-confluent flows in this case
  is \[\frac{2mk + 3}{mk + m + 2} = 2 - \frac{2m + 1}{mk + m + 2}
  \underset{m \to \infty}{\to} 2 - \frac{2}{k + 1}\,.\]

  We now show that the ratio achieved in this example is tight.
  Our ratio is similar to (and inspired by) the upper bound for the
  list-scheduling approximation
  ratio~\cite{graham_boundsmultiprocessing_1969} in machine scheduling.

\begin{theorem}
  The maximum ratio between the confluent flow and non-confluent flow
  in any grid polygon \(P\) is \(2 - (2 / (k + 1))\).
\end{theorem}

\begin{myproof}
  Let the ratio between the confluent flow and the non-confluent flow
  for a given polygon \(P\)
  be known as \(R_P\).  When calculating \(R_P\), we assume that the
  confluent and non-confluent flows are calculated optimally for
  \(P\).

  We begin by observing that by reducing the size of the smallest
  bottleneck in a polygon \(P\) can only increase \(R_P\).  Suppose we
  have polygons \(P_a\) and \(P_b\), where \(P_a\) has a minimum
  bottleneck size of at least 2, and \(P_b\) is the same as \(P_a\),
  except that one grid square has been removed from the minimum
  bottleneck.   The number of exit classes on one side of the smallest
  bottleneck in the confluent case can only decrease in \(P_b\)
  relative to \(P_a\), while in the
  non-confluent case, they may stay the same.  Thus, it is possible
  that there exists an exit in the non-confluent solution that is used
  for longer than in the confluent solution.  This implies that the
  number of steps that is required to evacuate the building in the
  confluent setting increases faster than the number of steps required
  under the non-confluent setting.  Therefore, \(R_{P_a}\)
  is at least as large as \(R_{P_b}\), and might be larger.

  This implies that \(R_P\) is maximized when the size of the minimum
  bottleneck is minimized, so we can assume that the size of the
  minimum bottleneck is 1.  We call the subpolygons on either side of
  the bottleneck \(P_1\) and \(P_2\).

  Furthermore, we can easily see that increasing the difference between the
  number of exits in \(P_1\) and \(P_2\) can only increase \(R_P\).
  This is because the number of people that must go
  through the bottleneck that separates \(P_1\) and \(P_2\) can only
  be increased by increasing this difference.  Therefore, we can
  assume that all the exits are on one side of the bottleneck between
  \(P_1\) and \(P_2\).  Without loss of generality, assume that all
  exits are in \(P_1\).

  Given this setup, we attempt to construct \(P\) so that as
  many exits as possible are used during as many time steps as
  possible in the non-confluent case.  This implies that there must be
  some source of people in \(P_1\).  This is because only one person
  can go through the bottleneck between \(P_1\) and \(P_2\) per time
  step.  Therefore, at most one person from \(P_2\) can reach an exit
  per time step.  However, by creating a supply of people in \(P_1\),
  we allow the people from \(P_2\) to queue in front of the exits.

  So that the people from \(P_2\) can queue in front of the exits, the
  route taken by the people in \(P_1\) from the supply to the exit
  must not interfere with the paths of the people from \(P_2\) to the
  exits.  This means that the number of people in \(P_1\) must be
  split and distributed to each exit.

  Therefore, \(P\) has the form sketched in Figure \ref{fig:max-ratio-shape}(b).
  There is a bottleneck \(B\).  The number of people behind \(B\) is
  \(x\), the amount of space for these people to queue in is \(y\),
  and the supply of people for each exit \(e_i\) is \(s_i\).

  In both the confluent and non-confluent solutions, it takes time
  \(2y / k\) to remove the \(y\) people in the queueing area.  In the
  non-confluent solution, it is necessary that this is the first step
  performed.  After this is done, the people from behind the
  bottleneck begin entering the queueing area.  The people must
  therefore take turns exiting from the queueing area and exiting from
  the supplies that are attached to each exit.  This implies that
  \(y\) is as small as possible (while satisfying \(y \ge 2k\)) and
  that \(s_i \ge x\) for all \(1 \le i \le k\).  Having different
  values of \(s_i\) provides no advantage, so we assume that the value
  of \(s_i\) is some value \(s_x\) for all \(i\).

  The ratio between the confluent flow and non-confluent flow is thus
  \[\frac{2y / k + 2s_x + 2x}{2y / k + 2s_x + 2x / k}\] which is
  maximized according to our constraints when \(s_x = x\) and when \(y
  = 2k\).  This gives a ratio of \[\frac{2x + 2k / k}{x + x / k + 2k /
  k} = \frac{2x + 2}{x + x / k + 2} = 2 - \frac{2x + 2k}{xk + x + 2k}
  \underset{x \to \infty}{\to} 2 - \frac{2}{k + 1}\] which is the claimed result.
\end{myproof}
  
\section{Conclusions}
\label{sec-4}

  \label{sec:conclusions}

  We have discussed evacuations in grid polygons.  We first showed
  that finding evacuations with confluent flows in polygons with holes
  is hard, even for polygons with only two exits.  We then looked at
  algorithms to find evacuations with confluent flows.
Finally, we
  showed that, while the difference between confluent and
  non-confluent flows is potentially significant, it is bounded.

  Our work raises some questions that require further study.  
  For simple polygons,
  there is evidence that a constant number of exits allows
  strongly polynomial solutions, even though some of the technical
  details are complicated.  
  What is the complexity of finding an evacuation plan with a
  confluent flow when the number of exits is not constant?  Next, can
  we find an fixed-parameter-tractable algorithm to find the confluent
  evacuation of polygons?  Finally, can we find a polynomial algorithm
  that gives the optimal evacuation using non-confluent flows?  Note
  that it is not even clear that the output size of such an algorithm
  is always polynomial.
  
\section*{Acknowledgments}
We thank Estie Arkin, Michael Bender, Joe Mitchell, and Martin Skutella
for helpful discussions; Martin Skutella is also part of ADVEST.

\bibliographystyle{abbrv}
\bibliography{evacuation}

\begin{thebibliography}{10}

\bibitem{bast_area_2000}
H.~Bast and S.~Hert.
\newblock The area partitioning problem.
\newblock In {\em Proc. 12th Can. Conf. Comput. Geom. {(CCCG} '00)}, pages
  163---171, Fredericton, NB, Canada, 2000.

\bibitem{bs-eafms-09}
N.~Baumann and M.~Skutella.
\newblock Earliest arrival flows with multiple sources.
\newblock {\em Math. Oper. Res.}, 34(2):499--512, 2009.
\newblock Journal version of 2006 FOCS article ``Solving evacuation problems
  efficiently''.

\bibitem{becker_max-min_1998}
R.~Becker, I.~Lari, M.~Lucertini, and B.~Simeone.
\newblock Max-min partitioning of grid graphs into connected components.
\newblock {\em Networks}, 32(2):115--125, 1998.

\bibitem{chazelle_triangulatingsimple_1991}
B.~Chazelle.
\newblock Triangulating a simple polygon in linear time.
\newblock {\em Discrete and Computational Geometry}, 6(5):485---524, 1991.

\bibitem{garey_computers_1979}
M.~R. Garey and D.~S. Johnson.
\newblock {\em Computers and Intractability: A Guide to the Theory of
  {NP-Completeness}}.
\newblock W. H. Freeman, 1979.

\bibitem{graham_boundsmultiprocessing_1969}
R.~L. Graham.
\newblock Bounds on multiprocessing timing anomalies.
\newblock {\em {SIAM} Journal on Applied Mathematics}, 17:416---429, 1969.

\bibitem{gyri_division_1978}
E.~Györi.
\newblock On division of graphs to connected subgraphs.
\newblock In {\em Combinatorics}, pages 485--494, Keszthely, 1978.

\bibitem{lovsz_homology_1977}
L.~Lovász.
\newblock A homology theory for spanning trees of a graph.
\newblock {\em Acta Mathematica Hungarica}, 30(3):241--251, 1977.

\bibitem{lumelsky_polygon_1998}
V.~Lumelsky.
\newblock Polygon area decomposition for multiple-robot workspace division.
\newblock {\em Int. Journal of Computational Geometry and Applications},
  8(4):437--466, 1998.

\bibitem{ma_$ok^2n^2$_1994}
J.~Ma and S.~Ma.
\newblock An {{$O(k^2n^2)$}} algorithm to find a {$k$}-partition in a
  {$k$}-connected graph.
\newblock {\em Journal of Computer Science and Technology}, 9(1):86--91, 1994.

\bibitem{moore_hard_2001}
C.~Moore and J.~Robson.
\newblock Hard tiling problems with simple tiles.
\newblock {\em Discrete and Computational Geometry}, 26(4):573--590, Dec. 2001.

\bibitem{salgado_approximation_2004}
L.~R. Salgado and Y.~Wakabayashi.
\newblock Approximation results on balanced connected partitions of graphs.
\newblock {\em Electr. Notes in Discrete Mathematics}, 18:207--212, Dec. 2004.

\bibitem{skutella_introduction_2009}
M.~Skutella.
\newblock An introduction to network flows over time.
\newblock In {\em Research Trends in Combinatorial Optimization}, pages
  451--482. {Springer-Verlag}, 2009.

\bibitem{suzuki_linear_1990}
H.~Suzuki, N.~Takahashi, and T.~Nishizeki.
\newblock A linear algorithm for bipartition of biconnected graphs.
\newblock {\em Information Processing Letters}, 33(5):227--231, 1990.

\bibitem{toussaint_solving_1983}
G.~Toussaint.
\newblock Solving geometric problems with the rotating calipers.
\newblock {\em In Proc. {IEEE} {MELECON} ’83}, pages 10---02, 1983.

\end{thebibliography}

\end{document}